\documentclass{article}
\usepackage{graphicx}
\usepackage{algorithm}
\usepackage{algpseudocode}
\usepackage{xcolor}
\usepackage[utf8]{inputenc}
\usepackage{blindtext}
\usepackage{geometry}
\usepackage{amsmath,mathtools}
\usepackage[
backend=biber,
style=alphabetic,
sorting=ynt
]{biblatex}

\title{\vspace{-3.0cm}Deterministic Algorithmic Approaches to Solve Generalised Wordle.}
\author{Aditya Lahiri, , Shivaank Agarwal, Vignesh Nanda Kumar}
\author{Aditya Lahiri \\{adlahiri@ucsd.edu} 
   \and Naigam Shah \\  {n7shah@ucsd.edu} 
   \and Shivaank Agarwal \\
   {s2agarwal@ucsd.edu} 
   \and Vignesh Nandakumar  \\ {vnandakumar@ucsd.edu}}

\begin{document}
\date{\vspace{-5ex}}
\maketitle

\section{Introduction}

Wordle is a single-player word-based game where the objective is to guess the 5-letter word in a maximum of 6 tries. The game was released to the public in October 2021 and has since gained popularity with people competing against each other to maintain daily streaks and guess the word in a minimum number of tries. There have been works using probabilistic and reinforcement learning based approaches to solve the game \cite{anderson2022finding},\cite{bhambri2022reinforcement}. Our work aims to formulate and analyze deterministic algorithms that can solve the game and minimize the number of turns required to guess the word and do so for any generalized setting of the game.\\
As a simplifying assumption, for our analysis of all the algorithms we present, we assume that all letters will be unique in any word which is part of our vocabulary. We propose two algorithms to play Wordle - one a greedy\cite{vince2002framework} based approach, and other based on Cliques. The Greedy approach is applicable for both hard and easy modes of Wordle, while the Clique formation\cite{balas1986finding} based approach only works on the Easy mode. We present our analysis on both approaches one by one, next.

\section{Problem Statement}
Given the vocabulary consisting of fixed-length words from the English dictionary and an unseen hidden word (which also belongs to the vocabulary), our aim is to guess the hidden word. Both the algorithms we present take as input the vocabulary and the previous guesses made, and hence can also be used to find the best next move if a player decides to use the algorithm after a few moves and not at the beginning of the game.
\subsection{Input}
\begin{itemize}
    \item Number of letters in the word: $l$ (varies from 3 to 9)
    \item Vocabulary $Vocab$ of size $n$, consisting of all $l$ letter words
    \item Unseen hidden word $h$, of size $l$
\end{itemize}
\subsection{Output}
\begin{itemize}
    \item Guess word $w\_g$ at each round, where $w\_g \in Vocab$
    \item Number of guesses $m$ it takes to get to the hidden word $h$
\end{itemize}

\section{Board State}
At any point in the game, our board consists of a $ l \times k$ grid where $l$ denotes the number of guesses taken till now ($0 \leq l \leq m$) and $k$ denotes the number of letters in each word. Formally we define our board $B$ as,
$$ B[i][j] : \;\;  1 \leq i \leq m,\;\; 1\leq j \leq k$$
Where each $B[i][j]$ is a tuple consisting of a letter and a colour, defined as follows:
$$ B[i][j] = (letter,color)$$
$$ letter \in [A-Z] $$
$$ color \in (green, yellow, gray) $$
Since each guess should be a valid word in the vocabulary set,
$$ B[i] \in D_k \;\; \forall i$$
The $color$ variable in each $B[i][j]$ depends on whether the letter guessed in at $i,j$ belongs to the hidden word $h$ and whether it is in the correct position or not. $green$ denotes that the letter belongs to the word and is in the correct position, $yellow$ denotes the letter belongs to the hidden word but is not in the correct position and gray denotes that the letter does not belong to the hidden word. We formally define this as:
$$ \text{If} \;\; B[i][j].letter =h[j] \;\; \text{then} \;\; B[i][j].color = green$$
$$ \text{If} \;\; B[i][j].letter \in h \;\; \text{but} \;\; B[i][j].letter \neq h[j] \;\; \text{then} \;\; B[i][j].color = yellow$$
$$ \text{If} \;\; B[i][j].letter \notin h \;\; \text{then} \;\; B[i][j].color = gray$$
\section{Defining a Player Move}

Assuming that we have the entire input set as defined above, and the board $B$ already has $i$ rows, each denoting color encoding of user's previously $i$ guessed words. Now, for each of the two modes of the game, namely easy and hard modes, the user will input their $(i+1)^{th}$ word of consisting of k-letters. Depending on which mode the game is being played at, the game will check the validity of the word. If its a valid word then the game will reveal the colors of all the $k$-letters of the $(i+1)^{th}$ word w.r.t the hidden word $h$, otherwise the game asks the user to try again with a different word. Note that in case of \textit{Hard mode}, if a letter has been guessed correctly and in the right position in one of the earlier guesses, the letter has to be a part of the guess in the same location in all the subsequent chances. Similarly, if a letter has been guessed correctly but is in the wrong position, it has to be guessed again in at least one of the positions in all the subsequent guesses. In \textit{Easy mode}, we are not forced to use a letter identified previously as correct in any of the subsequent guesses. Formally we define this as:

\begin{enumerate}
    \item \textbf{Easy mode:} \\
    The user inputs their $(i+1)^{th}$ guess word $w$.

    \textbf{If} $w \in D_k$, \textbf{then} add a new row to board $B$ with row $B[i+1]$ denoting the colors of $k$-letters in $w$ as follows: \\
    \hspace*{10mm} \textbf{If}  $w[j]=h[j]$ \textbf{then} $B[i+1][j].color = green$\\
    \hspace*{10mm}  \textbf{If}  $w[j] \in h$ but $w[j]\neq h[j]$ \textbf{then} $B[i+1][j].color = yellow$\\
    \hspace*{10mm} \textbf{If}  $w[j] \notin h$ \textbf{then} $B[i+1][j].color = gray$

    $ \text{\textbf{Else}} \;\; \text{return invalid word, ask the user to try again}.$\\

    \item \textbf{Hard mode:} \\
    The user inputs their $(i+1)^{th}$ guess word $w$.

    \textbf{If} $i \geq 1$, $B[i][j].color = green$ and $B[i+1][j].letter \neq h[j]$ for any $j$ \textbf{then} return invalid word, ask user to try again.

    \textbf{Else If} $i \geq 1$ and for any $j'$, $B[i][j'].color = yellow$ and $B[i][j'].letter \neq B[i+1][j]$ for any $j$ \textbf{then} return invalid word, ask user to try again.

    \textbf{Else} add a new row to board $B$ with row $B[i+1]$ denoting the colors of $k$-letters in $w$ as follows: \\
    \hspace*{10mm} \textbf{If}  $w[j]=h[j]$ \textbf{then} $B[i+1][j].color = green$\\
    \hspace*{10mm}  \textbf{If}  $w[j] \in h$ but $w[j]\neq h[j]$ \textbf{then} $B[i+1][j].color = yellow$\\
    \hspace*{10mm} \textbf{If}  $w[j] \notin h$ \textbf{then} $B[i+1][j].color = gray$
\end{enumerate}
\section{Objective function}
There are $4$ possible objectives that must be optimized for while solving the game of Wordle.\\
\begin{itemize}
    \item Output parameter $is\_solved$ is $True$. In other words, solve the game such that $num\_moves$ is $\le m$, and $B[num\_moves].letters=h$
    \item $min$ $(num\_moves)$. In addition to simply solving the game, minimize the number of guesses it takes to solve it.
    \item Best first move: Find the first guess word when $l=0$, i.e. when board $B$ is empty, such that, this word leads to $min$ $(num\_moves)$, average minimum number of moves needed to solve the game.
    \item Best next move (Given current board state): Given state of Board $B$, and current move $l \ge1$, find best next word $w \in D_k$ to guess such that,  $w$ leads to $min$ $(num\_moves)$ i.e. minimum number of moves needed to solve the game.

\end{itemize}

\section{Algorithms}

\subsection{Greedy Algorithm}
The first algorithm we present to solve wordle is the greedy approach. In this algorithm, we prune the vocabulary set based on the current guess and the outcome of the guess. The guess word is decided based on the greedy strategy of the choice which will have the lowest number of remaining words in the pruned vocabulary, in the worst case. 

\subsubsection{Algorithm and description}
In the greedy algorithm, we use 3 helper functions, described as follows:
\begin{itemize}
    \item \textbf{PlayWordle($Vocab,h$)} - Takes as input the entire vocabulary and the hidden word to be guessed, and processes each guess by giving a pattern of gray, yellow, and green letters based on the guess word and the hidden word.
    \item \textbf{GetPattern($w\_g,w\_h$)} - Returns the pattern that would be obtained if $w\_g$ is guessed when $w\_h$ is the hidden word.
    \item \textbf{TrimVocab($Voacb,w\_g,pattern$)} - Returns a trimmed vocabulary based on the $pattern$ obtained when $w\_g$ is guessed.
\end{itemize}
Algorithm \ref{algorithm 1 helper} shows the pseudo-code for all the above 3 helper functions.
\begin{algorithm}
\caption{Greedy algorithm helper functions}\label{algorithm 1 helper}
\begin{algorithmic}[1]

\Function{PlayWordle}{$Vocab,h$}
    \While{$len(Vocab)>1$} 
        \State $guess\_word \gets Greedy(Vocab)$

        \State $Vocab \gets TrimVocab(Vocab, guessword, pattern$)
    \EndWhile
\EndFunction 

\Function{GetPattern}{$w\_g,w\_h$}
    \State $pattern$ = [ ]
    \For{$index \in l$}
    \Comment{Where $l$ is the length of every word in $Voacb$}
    \If{$w\_g[index] = w\_h[index]$} 
    \State $pattern.append('green')$ 
    \Comment{correct letter in correct position} 
    \ElsIf{$w\_g[index] \in w\_h$} 
    \State$pattern.append('yellow')$ 
    \Comment{correct letter in incorrect position}
    \Else 
    \State$pattern.append('grey')$ 
    \Comment{grey if incorrect letter}
    \EndIf
    \EndFor
    \State \textbf{return} $pattern$
\EndFunction \\

\Function{TrimVocab}{$Vocab,w\_g,pattern$}
    \State $trimmed\_Vocab$ = [ ]
    \For{$word \in Vocab$}
        \If{GetPattern($w\_g,word$) = $pattern$} 
            \State $trimmed\_Vocab$.append($word$) \\
        \Comment{Appending word if pattern matches} 
        \EndIf
    \EndFor
    \State \textbf{return} $trimmed\_Vocab$
\EndFunction \\

\end{algorithmic}
\end{algorithm}

Using the above helper functions we describe our greedy algorithm. The pseudo-code for the greedy approach is given in Algorithm \ref{greedy algorithm}. Given vocabulary $Vocab$, we choose a word to guess ($w\_g$) and use that as our next guess in the game. After making that guess we will obtain a pattern using which we can prune out all the words which cannot be the hidden word, based on the pattern we obtain. Our aim is to choose a word to guess which will prune out the maximum words in the vocabulary, in the worst case. \\
We do this by iterating over all possible words ($w\_g$) in our $vocab$ as they act as our candidate words to be guessed (line 4). For each candidate word $w\_g$, we iterate over all possible words ($w\_h$) in the entire vocabulary as they act as the possible hidden words of the game (line 7) . For each $w\_h$, we obtain the pattern of green,yellow and gray letters we would get if we guessed $w\_g$ when the hidden word was $w\_h$ (line 8). We keep a track of the number of hidden words which belonged to all the patterns. After iterating over all $w\_h$ for one particular $w\_g$, we store the highest number of words which belonged to any pattern for the $w\_g$ (line 15). This represents the worst case number of words which would be in our pruned vocabulary if $w\_g$ was guessed. This process is repeated over all possible $w\_g$'s in the dictionary. We return the $w\_g$ which has the lowest worst case number of remaining words (line 19). After we make a guess, we prune our vocabulary based on the guess we made and the pattern we obtained from the game, and continue the process of choosing a guess word based on the pruned vocabulary.This process continues till the size of the vocabulary is 1 or the word is guessed correctly.

\begin{algorithm}[H]
\caption{Algorithm}\label{greedy algorithm}
\begin{algorithmic}
\Function{Greedy}{$Vocab$}
\State $worst\_case\_words \gets [0,0,...0]$ \\
\hspace{1mm} \Comment{Zero array of size $Vocab$ containing worst case scenario for every word}
\For{$w\_g \in Vocab$}
    \State $patterns \gets [0,0,...0]$ \\
    \hspace{1mm} \Comment{Hash map of size $3^l$ containing number words belonging to each pattern}
    \For{$w\_h \in Vocab$}
        \State $cur\_pattern \gets $ GetPattern($w\_g,w\_h$)\\
        \Comment{Get pattern if $w\_g$ was guessed when $w\_h$ is the hidden word}\\
        \State $patterns[cur\_pattern] += 1$ \\
        \Comment{Increment index of corresponding pattern}\\
    \EndFor
    \State $worst\_case\_words[w_g] \gets max(patterns)$ \\
    \Comment{Storing worst case possible of current word}\\
\EndFor \\
\Return $Vocab[argmin(worst\_case\_words)]$ \\
\Comment{Returning index of guess word having the lowest worst case remaining words}\\
\EndFunction \\

\end{algorithmic}
\end{algorithm}

\subsubsection{Correctness}
\textbf{\textit{Claim 1:}} Given a guess word $w\_g$, all words belong to one of the possible $3^l$ patterns. \\
\textbf{\textit{Proof:}} Given a guess word $w\_g$, and any word $w$ from the vocabulary, both of these are of the same number of letters - $l$ as are all the words in the vocabulary. Also both $w$ and $w\_g$ contain only one of the 26 alphabets. Hence each letter in $w$ can either be in $w\_g$ at the same index (green), or at some other index (yellow) or not  be in $w\_g$ (gray). Hence corresponding to each of the $l$ positions in $w$ we would get 1 of the 3 possible colors from $\{green,\,yellow,\,gray\}$. Therefore $w$ belongs to one the possible $3^l$ patterns for any $w$ in the vocabulary.

\noindent \newline \textbf{\textit{Claim 2:}} All possible hidden words $w\_h$ that belong to the same pattern, represent the set of candidates for the hidden word, or are pruned completely. \\
\textbf{\textit{Proof:}} After we guess the word $w\_g$, the game returns a pattern of green,yellow and gray. If a word in the vocabulary $w$ gives the same pattern on computing GetPattern($w\_g,w$), there is no way to distinguish it from the actual hidden word in the game. On the other hand if a word $w$ in the vocabulary gives a different pattern than the one returned by the game, it means that it has at least one letter which is different from that of the hidden word and hence is not a possible hidden word of the game. We pruned our all the possible words which do not give the same pattern as given by the game, and hence our pruned vocabulary contains only words which are possible candidates for the hidden word.

\noindent \newline \textbf{\textit{Claim 3:}} The greedy algorithm terminates after a finite number of guesses.
\textbf{\textit{Proof:}} The only possibility that the algorithm continues indefinitely is if at any round all words in the $vocab$ remain the same, i.e. no word is pruned. This is not possible as for every word in the current $vocab$, there will be only one word belonging to the pattern $green^1,green^2,...green^l$ which would be itself. Hence when a word is guessed, it either will be correct (hidden word) or be pruned out. So $vocab$ size will reduce by at least one word each round.\\

Using \textit{Claim 1} we know that the real hidden word belongs to one of the $3^l$ patterns as it is present in the vocabulary. Using \textit{Claim 2} we know that the hidden word will be present in the pruned vocabulary for all the rounds of the game, as it is always a possible candidate. We terminate our algorithm when we guess correctly or the $vocab$ size is 1. Using \textit{Claim 3} we know that vocab size decreases each round so the game will terminate. Hence we know that the hidden word is always part of the current $vocab$ and $vocab$ size decreases so we will eventually guess the hidden word correctly.

\subsubsection{Time Complexity}
The total time complexity of the greedy algorithm can be divided into 2 parts
\begin{itemize}
    \item Time taken in each round
    \item Total number of rounds in the worst case
\end{itemize}
The input to our algorithm is the vocabulary of size $n$ and the number of letters in each word which is $l$. We also assume each letter belongs to one of the 26 alphabets of the English dictionary. We define our complexity based on these 3 parameters.
\subsubsection{Time complexity for each round}
Time complexity for each component in 1 round:
\begin{itemize}
    \item Initializing $worst\_case\_words$ array of size $vocab$ = $O(n)$
    \item Outer loop runs $O(n)$ times
    \item Inner loop runs $O(n)$ times
    \item For each inner loop:
    \begin{itemize}
        \item GetPattern() does $l$ comparisons = $O(l)$
        \item incrementing $patters$ involves finding the position in hash map and incrementing it by 1 which is $O(1)$
    \end{itemize}
    \item Instead of finding maximum of all patterns as mentioned in the pseudo-code, we keep a track of maximum at each increment in the inner loop making it $O(1)$ at each iteration.
    \item Finding minimum of $worst\_case\_words$ takes operations equivalent to number of words which is $O(n)$
    \item For pruning vocabulary, we can go through each word once and compare its pattern to what we should get, making it $O(nl)$
\end{itemize}
Overall complexity of each round hence is :
$$ O(n) + O(n*n*(l+1+1)) + O(n) + O(nl)= O(n^2l)$$

\subsubsection{Total number of rounds in the worst case}
Since we do not have a defined structure for the number of letters that occur in each position in the English dictionary, as it varies with $l$ (length of the word), we assume that every combination of $l$ letters is a valid word in our vocabulary.\\
\noindent \newline \textbf{\textit{Claim 1:}} Vocabulary pruned by obtaining a pattern with 2 green/yellow instances will always have fewer words than vocabulary pruned by obtaining by a pattern with 1 green/yellow instances. \\
\textbf{\textit{Proof:}} If the game returns a pattern with a 2 green instances, we prune all the words in our vocabulary which do not have those 2 instances. So if the game returns a pattern with 1 green instance, the vocabulary obtained by pruning will be a super-set of the previous case. This is because it will contain all words which have 2 green instances along with words that have only one green instance. A similar argument holds for the case of yellow instances and can be generalized for 3,4,..l green/yellow instances. \\
\noindent \textbf{\textit{Claim 2:}} Number of possible words left in the vocabulary if the pattern returned by the game is all gray, 1 yellow (rest gray), and 1 green (rest gray) is ${^{26}P_{l}}$, ${^{26-l}P_{l-1}} \times (l-1)$, and ${^{26-l}P_{l-1}}$ respectively, where $P$ is the permutation symbol.\\
\textbf{\textit{Proof:}} We consider each case individually:
\begin{itemize}
    \item \textbf{Case 1: All gray letters - } In this case our pattern obtained is all gray, indicating none of the letters w from the word we guessed belong to the hidden word. This means that we can prune our vocabulary to only contain permutations of the rest of the 26-k alphabets, leading to a new vocab size of $^{26-l}P_{l}$.
    \item \textbf{Case 2: 1 yellow rest gray - } In this case we know that one letter belong to the hidden word but in some other position and the rest of the l-1 gray letters do not belong. So for the rest of the k-1 letters, we can permute over 26-(yellow alphabet)-(l-1 gray alphabets) = 26-l alphabets since we know that all alphabets in a word occur only once. The yellow letter can be present in any of the remaining l-1 positions, leading to a new vocab size of $^{26-l}P_{l-1} \times (l-1)$.
    \item \textbf{Case 2: 1 green rest gray - } In this case we know that one letter belong to the hidden word in the same position and the rest of the l-1 gray letters do not belong. So for the rest of the l-1 letters we can permute over 26-(green alphabet)-(l-1 gray alphabets) = 26-l alphabets, leading to a new vocab size of $^{26-l}P_{l-1}$.
\end{itemize}

\noindent \newline \textbf{\textit{Claim 3:}} Having all gray letters gives the largest possible remaining vocabulary size for all words of length less than 10. \\
\textbf{\textit{Proof:}} Using claim 1 we know that having multiple yellow/green letters prunes more words in the cocab as compared to having only a single yellow/green letter. Using claim 2, we know that having 1 green alphabet leads to a vocab size of $^{26-l}P_{l-1}$ which is less than the vocab size of 1 yellow and all gray given by $^{26-l}P_{l-1} \times (l-1)$. Now we compare the vocab sizes of all gray versus 1 yellow and the rest gray:
\begin{center}
    Vocab size of all yellow < Vocab size of all gray
\end{center}
$$ ^{26-l}P_{l-1} \times (l-1) < ^{26-l}P_{l}$$
$$ \displaystyle \frac{(26-l)!}{(27-2l)!} \times (l-1) < \frac{(26-l)!}{(26-2l)!}$$
$$ \displaystyle \frac{1}{(27-2l)} \times (l-1) < \frac{1}{1}$$
$$ 3l < 28 \implies l < 10$$
\begin{center}
    (As l is the length of the word, an integer)
\end{center}

Using claims 1-3 we know that having a response pattern of gray letters will lead to the largest size of the pruned vocabulary, and will reduce the vocabulary size to a combination of $n-l$ alphabets when $n$ is the size of the previous vocabulary and $l$ is the length of the word. So we can have all gray letters a maximum of $\lceil{\frac{26}{l}} \rceil$ times, as after that we are bound to know all letters which will occur in our hidden word, but not necessarily their positions. \\
Once we know all the alphabets that occur in the hidden word, we use algorithm \label{attr:check_all_anagrams} to reach our hidden word in $l$ guesses. At the high level, the algorithm shifts the letters of our guessed word by 1 position each turn until the letters are in the correct position. So after $l$ turns, each letter has had the chance of going through all positions and hence $l$ is the maximum number of turns it can take for the algorithm to find the hidden word. \\
The overall number of rounds in the worst case is:
$$ \displaystyle  \lceil{\frac{26}{l}} \rceil + l$$

Therefore overall complexity for the greedy algorithm is:
\begin{center}
    (Complexity of each round) $\times$ (No. of rounds in worst case)
\end{center}
$$ = (O(n^2l)) \times (\lceil{\frac{26}{l}} \rceil + l) = \boldsymbol{O(n^2l^2)} $$.
\subsection{Clique Based Algorithm}

\subsubsection{Helper Functions and Data Structures Overview}

\begin{enumerate}
    \item Wordle Tracker (\emph{Class})
    \item Form Graph (Input : $wordle\_tracker$, $bool$:$hard$)
    \Return $adj\_list$,$edge\_exists$
    \item Find k clique(Input : $adj\_list$, $int$:$k$)
     \Return $cliques$
    \item Process cliques(Input : $wordle\_tracker$, $List$:$cliques$)
     \Return $wordle\_tracker$
    \item Check all anagrams(Input : $wordle\_tracker$, $string$:$hidden\_word$)
    \item Guess remaining words(Input : $wordle\_tracker$, $string$:$hidden\_word$)
\end{enumerate}

\subsubsection{Play Wordle Algorithm using Clique Finding}
We use the PlayWordle Using Clique algorithm[\ref{alg:play_wordle_clique}] to solve the game of Wordle using cliques. This uses the helper functions defined above to help us play the game. We start the game by initializing an object of the Wordle Tracker[\ref{attr:wordle_tracker}] class. The game is then played until all letters are found. To do this, we construct a graph based on current state of Wordle Tracker using the  Form Graph[\ref{attr:form_graph}] function. Given this graph, we find the most informative clique using the FindKClique[\ref{attr:find_k_clique}] and ProcessCliques[\ref{attr:process_clique}] functions. if all letters of the hidden word are found, we run a check on all possible anagrams using CheckAllAnagrams[\ref{attr:check_all_anagrams}] function. If we are unable to find all letters of the word, we guess the remaining possible words in the vocabulary using the GuessRemainingWords [\ref{attr:guess_remaining_words}] function. After this, we have the updated WordleTracker object which provides us with all the words we guessed and also our final guess which ends the game.\\
In the next sections, we describe the algorithm, working, runtime, and correctness of all of these components. Finally, in [\ref{attr:overall_correct_time}] we analyse the overall runtime and correctness of our clique based algorithm to play Wordle.
\begin{algorithm}
\caption{Play Wordle Algorithm using Clique}\label{alg:play_wordle_clique}
\begin{algorithmic}[1]

\Function{PlayWordle}{$vocab, hidden\_word$}
    \State $word\_length \gets len(hidden\_word)$
    \State $ALPHABET\_SIZE \gets 26$
    \State $wordle\_tracker \gets WordleTracker(vocab)$
    \Comment{Initializing wordle tracker}
    \While{$len(wordle\_tracker.letters\_found) != word\_length$}
        \State $word\_graph, edge\_exists \gets FormGraph(wordle\_tracker)$
        \If{$edge\_exists == False$}
        \Comment{Not able to form any graph}
            \State $break$
        \EndIf
        \State $max\_clique\_size \gets ALPHABET\_SIZE//word\_length$
        \State $all\_cliques \gets FindKClique(word\_graph, k=max\_clique\_size)$
        \If{$len(all\_cliques)$}
        \Comment{If we are not able to get any clique}
            \State $break$
        \EndIf
        \State $wordle\_tracker \gets ProcessCliques(all\_cliques, wordle\_tracker, hidden\_word)$
    \EndWhile
    \If {$len(wordle\_tracker.letters\_found) != word\_length$}
        \State $wordle\_tracker \gets CheckAllAnagrams(wordle\_tracker, hidden\_word)$
    \Else
        \State $wordle\_tracker \gets GuessRemainingWords(wordle\_tracker, hidden\_word)$
    \EndIf
\EndFunction

\end{algorithmic}
\end{algorithm}

\subsection{Wordle Tracker}
\label{attr:wordle_tracker}
\subsubsection{Algorithm description}
As defined in the project formulation, we keep track of current board state with all the words guessed so far. The yellow (correct letter wrong position), and green (correct letter correct position) letters are tracked in separate data structures. We will go over those and also other helper functions defined. \\

\begin{enumerate}
    \item Vocab : Vocab is a list of all words that can be guessed.
    \item Unseen chars: Set of all unseen characters in the current board state. It will be initialized to all 26 letters before the first move since no word has been guessed yet and all letters are unseen. We need to track this because we use this to choose the best clique if we find multiple k cliques as part of our clique finding routine.
    \item Discarded words: Set of words that are definitely not the correct answer based on the current board state. Initially the set of discarded words is empty.
    \item Letters found: A set of letters containing all yellow and green letters.
    \item Letter positions: a hash map containing green letters and their corresponding positions.
    \item Words guessed: This is a list storing the words guessed so far.
    \item Grey letters: a set of grey letters (letters not present in the hidden word based on the guesses done so far).
\end{enumerate}

Now we will go over the helper functions we will use for the clique based wordle solver algorithm. All these helper functions will update the state of the board based on the guesses. 

\begin{enumerate}
    \item UpdateWordsGuessed: This function will just append the list of words guessed to the words guessed array.
    \item UpdateLettersFound: This function goes over the list of guessed words and for each word computes which letter will be green/yellow/grey by comparing it with the hidden word. If the position and letter match for with the corresponding hidden word position, then it is a green letter. Else if just the letter guessed is present in hidden word, then it is a yellow letter. Else, it will be a grey letter.
    \item UpdateUnseenChars: This function computes the set of unseen characters in the current board state. This is done by finding the set of seen letters by computing the union between all green, yellow and grey letters. Then unseen characters just the set difference between the current unseen characters and the set of seen characters.
    \item UpdateDiscardedWords: This function iterates over all words not in the current discarded set of words and checks if any word needs to be added to the discarded words set. If the word has any letter in common with set of grey words or if a yellow letter is not present in it or if a green letter is not present in the right position then we add the word to the set of discarded words.
\end{enumerate}

\subsubsection{Proof of Correctness}

\textbf{\textit{Claim:}} The wordle tracker state is updated correctly after each guess.

\textbf{\textit{Proof:}} We will be using weak induction to prove that all the helper functions defined update the wordle tracker state correctly.

Induction hypothesis: The wordle tracker state is updated correctly after k guesses.

Base case: Before any guess, when k = 0, the list of words guessed is empty, none of the letters are found and hence the green letter positions are also empty. The unseen characters set contains all 26 letters since no word has been guessed. The discarded words set is also empty since all words can be possible answers.

Induction hypothesis: The wordle tracker state is updated correctly after k guesses.

To prove: The wordle tracker state is updated correctly after k+1 guesses.

Firstly the words guessed function will just append 1 more guess after k+1 guesses and it will hold k+1 words. So it is updated correctly.

Now coming to the set of letters found. After k+1 guesses, if we find any new yellow or green letter which gets us closer to the hidden word, we will update the letters\_found and letter\_positions accordingly. We add to letter\_positions only if we find a green letter. And in case the letter is neither yellow nor green, it is added to the grey letters set. If a letter is not present in the hidden word, it will not be added to the letters found set. If a letter is present in the hidden word, it will never be added to the grey letters set. So, the letters found state is updated accordingly. 

Finally, to prove that the set of discarded words is updated correctly, whenever a new word is guessed we showed in the previous paragraph that the letters found and grey letters are updated correctly. So, we will use that to show discarded words are updated correctly. If a word contains any of the new grey letters added, it will be added to the discarded word set or if it already had any of the previously added grey letters, it would've already been added to the set because of induction hypothesis. Same reasoning can be used to show a word being added to discarded set if it does not contain any of the yellow/green letters found so far (or does not contain the green letter at the correct position). Any word which has all letters either yellow, green or unseen letters will not be added to the discarded set of words. Hence, we show that the discarded set of words is updated correctly.

Hence, we prove the induction claim that the wordle tracker state is updated correctly after k+1 guesses.

\begin{algorithm}
\caption{Helper Functions for Wordle Tracker}\label{alg:helper_func}
\begin{algorithmic}[1]
\Function{UpdateWordsGuessed}{$guessed\_words\_indices$} \\
    \Comment{Append List of guessed words to the words guessed array}
\EndFunction \\

\Function{UpdateLettersFound}{$guessed\_words\_indices, hidden\_word$}
    \State $guessed\_words \gets$ [words corresponding to indices from vocabulary]
    \For {$w \gets guessed\_words$}
        \For {$i \gets len(w)$}
            \If{$w[i] == hidden\_word[i]$}
                \State $letters\_found.add(w[i])$
                \State $letters\_positions[w[i]] = i$
            \ElsIf {$w[i] \in hidden\_word$}
                \State $letters\_found.add(w[i])$
            \Else
                \State $grey\_letters.add(w[i])$
            \EndIf
        \EndFor
    \EndFor
\EndFunction \\

\Function{UpdateUnseenChars}{}
    \State $all\_seen\_letters \gets letters\_found \cup grey\_letters $
    \State $unseen\_chars \gets unseen\_chars - all\_seen\_letters $
\EndFunction \\

\Function{UpdateDiscardedWords}{}
    \Comment Check every word not already in the discarded word set
    \For {$w \gets vocab$} 
        \If{$w \in discarded\_words$}
            \State $continue$
        \EndIf
        \State $common\_grey \gets set(w)\cap grey\_letters$
        \If {$len(common\_grey) > 0$}
            \State $discarded\_words.add(w)$
        \EndIf
        \For {$l \gets letters\_found$}
            \If{$l \notin w$}
                \State $discarded\_words.add(w)$
            \EndIf
            \If {$l \in letter\_positions$ and $ w[letter\_positions[l]] != l$}
                \State $discarded\_words.add(w)$
            \EndIf
        \EndFor
    \EndFor
\EndFunction

\end{algorithmic}
\end{algorithm}

\subsubsection{Time analysis}
Let k be the number of words sent to these functions. Let length of word be l. Let n be the number of words in the vocabulary.

UpdateWordsGuessed : O(k)

UpdateLettersFound : O(kl) There are k iterations of the loop and in each iteration we iterate over each letter of a word and update yellow, green or grey letters.

UpdateUnseenCharacters : O(1) Let size of alphabet be a. Union and difference operations can be done in constant time.  So, we can implement the sets using a fixed size boolean array setting True on elements seen and False on elements not seen. So, union operation will be an OR operation across all elements one by one and difference will also be O(a) = O(1). 

UpdateDiscardedWords: 
For each word, we first check if the word has anything common with grey letters : O(1)
For each word, we go over each letter found and check if the word should be discarded. O(l)

So, total complexity of the loop = n iterations * (l)
 = O(nl)

 Total complexity = $O(nl)$

\subsection{Graph Formation}
\label{attr:form_graph}

\begin{algorithm}
\caption{Form Graph Helper}\label{alg:form-graph-helper}
\begin{algorithmic}[1]
\Require $wordle\_tracker$: the Wordle tracker object, $hard$: boolean indicating whether to use hard or soft constraints
\Ensure $graph$: adjacency list representation of the graph, $edge\_exists$: boolean indicating whether an edge exists in the graph
\Function{FormGraphHelper}{$wordle\_tracker, hard$}
  \State $edge\_exists \gets \textbf{False}$
  \State $vocab \gets wordle\_tracker.vocab$
  \State $graph \gets [[] \text{ for } i \text{ in } \text{range}(\text{len}(vocab))]$
  \State $allowed\_unseen\_common \gets 0$
  \If{not $hard$}
    \State $allowed\_unseen\_common \gets 1$
  \EndIf
  \For{$i \textbf{ in } \text{tqdm}(\text{range}(\text{len}(vocab)))$}
    \State $w1 \gets vocab[i]$
    \If{$w1 \text{ in } wordle\_tracker.discarded\_words$}
      \State $\textbf{continue}$
    \EndIf
    \For{$j \textbf{ in } \text{range}(i+1, \text{len}(vocab))$}
      \State $w2 \gets vocab[j]$
      \State $words\_intersection \gets \text{set}(vocab[i]) \cap \text{set}(vocab[j])$
      \State $common\_unseen\_letters \gets words\_intersection \cap wordle\_tracker.unseen\_chars$
      \If{$\text{len}(common\_unseen\_letters) == allowed\_unseen\_common$}
        \State $graph[i].\textbf{append}(j)$
        \State $graph[j].\textbf{append}(i)$
        \State $edge\_exists \gets \textbf{True}$
      \EndIf
    \EndFor
  \EndFor
  \State \textbf{return} $graph$, $edge\_exists$
\EndFunction
\end{algorithmic}
\end{algorithm}

\begin{algorithm}
\caption{Form Graph}\label{alg:form-graph}
\begin{algorithmic}[1]
\Require $wordle\_tracker$: the Wordle tracker object
\Ensure $graph$: adjacency list representation of the graph, $edge\_exists$: boolean indicating whether an edge exists in the graph
\Function{FormGraph}{$wordle\_tracker$}
  \State $new\_graph, edge\_exists \gets \text{FormGraphHelper}(wordle\_tracker, hard=True)$
  \If{not $edge\_exists$}
    \State $new\_graph, edge\_exists \gets \text{FormGraphHelper}(wordle\_tracker, hard=False)$
  \EndIf
  \State \textbf{return} $new\_graph$, $edge\_exists$
\EndFunction
\end{algorithmic}
\end{algorithm}

"FormGraph" takes a WordleTracker object and calls the "FormGraphHelper" function with the hard parameter set to True. If there are no edges in the graph returned by "FormGraphHelper", the function calls "FormGraphHelper" function with the hard parameter set to "False" . We make this hard and soft graph creation different to allow for more chances of edge creation in case of there being no two words having all distinct letters between them in some specific vocabulary. The soft graph creation allows for some lax and permits 1 letter to be common. Note that this has nothing to do with hard and soft game modes of the wordle game itself.

The "FormGraphHelper"  function takes a WordleTracker object and a boolean hard, which is True by default to allow for no common letters. The function creates an empty graph with an adjacency list data structure. It then sets the number of allowed common unseen letters between two words based on the value of hard to be generalizable for if and when it is called for soft graph creation case. For each word in the vocabulary, it checks if the word has already been discarded by the WordleTracker object. If so, it skips the word and moves on to the next word. Otherwise, it compares the current word with every other word in the vocabulary that comes after it. If the two words have the required number of common unseen letters(0 for hard and 1 for soft), an edge is added between the two vertices in the graph. The edge is bidirectional, so it is added to the adjacency list of both vertices. The EdgeExists flag is set to True if any edges were added to the graph. Finally, the function returns the graph and the EdgeExists flag.

\subsubsection{\textbf{Time Complexity}}The time complexity of the graph creation algorithm depends on the number of words in the input wordleTracker object, denoted by n. The time complexity of the algorithm is $O(n^2)$, as it needs to compare every pair of words. If the graph does not have any edge added to it due to non zero number of common letters(hard graph creation case), then the graph will be created through soft case where 1 unseen characters is allowed to be common, and the added time complexity will be $O(n^2)$. So, worst case the algorithm to create graph will run for $O(n^2)+O(n^2)$ which is $O(n^2)$

\subsubsection{\textbf{Correctness}}

\textbf{Claim}: The Graph formed is correct, ie, any two words that do not share any common letters (hard) have a vertex between them and no extra edges exist, ie, edges between pair of words that have any common letter(hard) (or at max 1 common letter in soft)

\textbf{Proof}: We will prove this using a loop invariant. The loop invariant is : After i iterations of the outermost loop, all the edges for the first i words in the vocabulary are added correctly.

Base case: Before the loop begins i = 0. So no words are seen yet so no edges are added to the graph. The loop invariant holds true.

Induction hypothesis: After i iterations of the outermost loop, all the edges for the first i words in the vocabulary are added correctly.

To prove: After i + 1 iterations of the outermost loop, all the edges for the first i words in the vocabulary are added correctly.

In the $(i+1)^{th}$ iteration, we iterate over all words j from i+2 to end and check for each pair of words (vocab[i+1], vocab[j]) if an edge needs to be added between them or not. An edge is added only if the condition of no unseen characters common (for hard graph) or 1 unseen character common (for soft graph) is satisfied. Further, no extra edges, ie, edges between words that share any letters ($\ge$ 2) are added to the graph as well. So, after $(i+1)^{th}$ iteration, the graph formed will be correct till i+1 nodes.

So, using this loop invariant, after n iterations, we can say that the graph formed is correct.

\subsubsection{Finding Cliques}
\label{attr:find_k_clique}

\begin{algorithm}[H]
\label{alg:find_k_clique}
\caption{Finding K-Cliques in a Graph}
\begin{algorithmic}[1]
\Require{An adjacency list, ${adjList}$, and an integer, $k$.}
\Ensure{A list of all $k$-cliques in $adjList$.}
\Function{findKclique}{$adjList, k$}
\State $cliques \gets []$
\Procedure{backtrack}{$current_clique, neighbours$}
\If{length of $currentClique$ is equal to $k$}
\State append $currentClique$ to $cliques$

\Else
\State $lastVertex \gets$ last element of $currentClique$
\For{vertex in $neighbours$}
\State $newNeighbours \gets neighbours \cap$ set of vertices adjacent to $vertex$
\If{vertex $>$ $lastVertex$}
\State $nextClique \gets currentClique$ with $vertex$ appended
\State backtrack($nextClique$, $newNeighbours$)
\EndIf
\EndFor
\EndIf
\EndProcedure
\For{index, vertex in enumerate($adjList$)}
\State backtrack($[index]$, set of vertices adjacent to $vertex$)
\EndFor
\State \Return{$cliques$}
\EndFunction
\end{algorithmic}
\end{algorithm}

\begin{algorithm}[H]
\label{alg:find_clique}
\caption{Finding Cliques in a Graph}
\begin{algorithmic}[1]
\Require{An adjacency list, ${adjList}$, and an integer, $max\_clique\_size$.}
\Ensure{A list of all cliques in $adjList$.}
\Function{findClique}{$adjList, max\_clique\_size$}
\State $i \gets 0$
\While{$len(all\_cliques)==0 \And max\_clique\_size-i>=3$}
    \State $all\_cliques \gets {findKclique}({adjList, max\_clique\_size - i})$ 
    \State $i \gets i+1$
\EndWhile
\State \Return{$all\_cliques$}
\EndFunction
\end{algorithmic}
\end{algorithm}

The two algorithms "Finding Cliques in A Graph" and "Finding K-Cliques in a Graph" together help us to find the clique of 
 maximum size that is available in the graph. The "Finding Cliques in A Graph" function iteratively tries to find the clique with the highest possible value currently, equal to $k$ by calling "Finding K-Cliques in a Graph" for different values of $k$ one by one, until one call results in a non-empty list of cliques.
The algorithm "Finding K-Cliques in a Graph"  takes in an adjacency list and an integer k, and returns a list of all k-cliques in the given graph represented by the adjacency list. The function makes use of a recursive backtracking algorithm to find all the cliques of size k. The algorithm works by iterating over each vertex in the graph and starting a backtrack from that vertex with a clique of size 1. The backtracking function takes in the current clique and a set of neighbouring vertices, and checks whether the current clique is of size k. If it is, the current clique is added to the list of cliques. Otherwise, the backtracking function loops over each neighbouring vertex and creates a new clique by appending that vertex to the current clique if the vertex is greater than the last vertex in the current clique. The reason for checking if vertex is greater than lastVertex in the backtrack function is to avoid generating duplicate cliques and to thereby optimize the runtime of the algorithm. This is because, starting from some vertex $v_i$ and then choosing some vertex $v_j$ is the same as starting from $v_j$ and then choosing $v_i$. Therefore, we can impose an arbitrary ordering on the vertices and only consider vertices that come after the last vertex in the current clique. 
The new clique is then passed recursively to the backtracking function along with the set of neighbours of the new vertex. The function then returns the list of all k-cliques that were found in the graph. As an \textit{optimization step}, we store all the cliques that are found as a memoization dictionary for upcoming multiple rounds of the game (different hidden words), since the cliques do not change for a given vocabulary.

\subsubsection{\textbf{Time Complexity}}The time complexity of this "Finding Cliques in a Graph"  $\sum_{i=3}^{k} (\binom{n}{i})$ where n is the number of vertices in the graph and k is the size of the cliques being searched for. This is because we run the Find i cliques In A Graph starting from k and going down in the worst case till 3 clique, and the k clique formation function has a runtime of N choose k. 

Complexity = $\sum_{i=3}^{k} (\binom{n}{i})$ $\le$  $\sum_{i=0}^{k} (\binom{n}{i})$ = $O(2^n)$ (binomial expansion)

\subsubsection{\textbf{Correctness}}
\textbf{Claim} : The K clique for the given input graph is found correctly.
\\
\textbf{Proof :}We can use induction on the size of the cliques being searched for (k). \emph{Base Case:} For k=2, the function is simply finding all edges in the graph, which is trivial.

\emph{Induction Hypothesis: }Now, Assume that the function works correctly for all values of k less than or equal to some integer m. 

\emph{To Prove:} We want to prove that the function works correctly for k=m+1. There can be two cases-
\begin{itemize}
    \item No clique of size m+1 or greater exists - We then returns an empty list, which is correct.
    \item Cliques of size m+1 or greater exist - For each vertex in the graph, the function finds all cliques of size m+1 that contain that vertex. This is done by recursively exploring all the vertices that are adjacent to the current vertex and have not already been included in the clique.  
\end{itemize}
 By the induction hypothesis, the function works correctly for cliques of size m, so we can assume that when the function is called recursively with a smaller clique size, it correctly finds all cliques of that size containing the given vertex.
 Therefore, when the function is called recursively with a clique size of m+1, it correctly finds all cliques of that size containing the given vertex. Since this is done for every vertex in the graph, the function correctly finds all cliques of size m+1 in the graph.

Thus proves the correctness of FindKClique function. \\

\textbf{Claim: } The FindClique function returns the k cliques for maximum possible k.

We will use a loop invariant to prove this statement. The loop invariant is : If the loop runs for i iterations, the it has not found k cliques for any k > max\_clique\_size - i.

Base case: For i = 0, the loop has not run, so it has not run the FindKClique function so no k cliques are found for k > max\_clique\_size.

Induction hypothesis: If the loop runs for i iterations, the it has not found k cliques for any k > max\_clique\_size - i.

To prove: If the loop runs for i+1 iterations, the it has not found k cliques for any k > max\_clique\_size - i - 1.

We have shown the correctness of FindKClique previously. So if the loop runs for i+1 iterations, we are not able to find any cliques in the first i iterations (induction hypothesis). In the i+1 iteration, we call FindKClique and it we find even a single clique, the size of all\_cliques will be > 0 and we will break out of the loop. And if no cliques were found, then the iterations will continue. Hence, we show that if the loop runs for i+1 iterations, the it has not found k cliques for any k > max\_clique\_size - i - 1.

Using the loop invariant we can say that if the loop ran for n iterations, we will return all cliques of size max\_clique\_size - n.

\subsubsection{Processing the cliques}
\label{attr:process_clique}

\textbf{Algorithm description}

The FindKClique function will return all possible k-cliques in the graph we create. Now we need to choose the best clique out of all the cliques using which we will guess our words. The intuition for choosing the best clique is to choose the clique which gives the most information. But we cannot cheat and guess ahead. We need to do this with the information we know. The information we know before a guess is the unseen set of characters so far. So, we would want to choose a clique that maximizes the number of unseen characters.

We define the function ProcessCliques that takes in three parameters: list of cliques (subsets of vertices in a graph), current wordle tracker that tracks the state of the Wordle game, and the hidden word the word that the player is trying to guess in the game.

The function begins by initializing some variables: vocab is a set of valid words, best\_clique is set to None, and best\_unseen\_chars\_covered is set to None.

Next, the function loops through each clique in clique\_list. For each clique, it calculates the set of unseen characters that are covered by the clique by calling the GetUnseenCharsCovered function. If the length of this set is greater than the length of the best unseen chars found so far, then the current clique is considered the best one found so far, and best\_unseen\_chars\_covered variables is updated accordingly.

Finally, the function calls the UpdateWordleState function to update the wordle\_tracker object with the words guessed which are part of the clique, letters found,  set of unseen characters after guessing and the set of discarded words after guessing the words part of the clique. The updated wordle\_tracker object is returned as the output of the function.

\begin{algorithm}
\caption{Processing all k cliques}\label{alg:proc_clique}
\begin{algorithmic}[1]

\Function{GetUnseenCharsCovered}{$clique, wordle\_tracker$} \\
    \Comment{Clique is a list of nodes}
    \State $curr\_unseen\_chars \gets wordle\_tracker.unseen\_chars$
    \State $vocab \gets wordle\_tracker.vocab$
    \State $chars\_covered \gets set()$
    \For{$n \gets clique$}
        \State $chars\_covered \gets chars\_covered \cup (set(vocab[n]))$
    \EndFor
    \State $unseen\_chars\_covered \gets chars\_covered \cap curr\_unseen\_chars$ \\
    \Return $unseen\_chars\_covered$
\EndFunction \\

\Function{UpdateWordleState}{$clique, unseen\_chars, wordle\_tracker, hidden\_word$}
    \State $wordle\_tracker.UpdateWordsGuessed(clique)$
    \State $wordle\_tracker.UpdateLettersFound(clique, hidden\_word)$
    \State $wordle\_tracker.UpdateUnseenChars()$
    \State $wordle\_tracker.UpdateDiscardedWords()$ \\
    \Return $wordle\_tracker$
\EndFunction \\

\Function{FindBestClique}{$clique\_list, wordle\_tracker, hidden\_word$}
    \State $best\_clique \gets None $
    \State $best\_unseen\_chars \gets set() $
    \For {$clique \gets clique\_list$}
        \State $unseen\_chars\_covered = GetUnseenCharsCovered(clique, wordle\_tracker)$
        \If{$ len(unseen\_chars\_covered) > len(best\_unseen\_chars) $}
            \State $best\_clique \gets clique$
            \State $best\_unseen\_chars \gets unseen\_chars\_covered$
        \EndIf
    \EndFor
    \State $new\_wordle\_tracker = UpdateWordleState(best\_clique, best\_unseen\_chars,$ \\ 
    $wordle\_tracker, hidden\_word)$ \\
    \Return $new\_wordle\_tracker$
\EndFunction

\end{algorithmic}
\end{algorithm}

\textbf{Proof of Correctness}

\textbf{\textit{Claim:}} Given a k-clique and the current state of wordle tracker, GetUnseenCharsCovered computes the unseen characters covered by the clique correctly.

\textbf{\textit{Proof}}: If a letter is previously unseen and covered by the clique it will be present in the unseen\_chars\_covered set. We add the the letters covered by the clique by computing the unions of all the letters in all the words part of the clique so the letter will be part of letters covered. Then the letter has to be previously unseen and now this clique had a word containing that letter. So, by computing intersection, we will correctly add the letter to the unseen\_chars\_covered set.

If the a letter is previously seen or not covered by the clique, it will not be added to the unseen\_chars\_covered set. If the letter is previously seen, it will not be part of curr\_unseen\_chars set. And if the letter is not covered by the clique, it will not be in the chars\_covered set.

\textbf{\textit{Claim: }} Loop invariant: After i iterations in the FindBestClique function, the loop keeps track of best clique so far in terms of the maximum unseen characters covered by the clique.

\textbf{\textit{Proof: }} Will use induction to prove this claim.

Base case: Before the loop starts the best clique is set to None i.e. since 0 iterations have happened, no clique is the best clique.

Induction hypothesis: After i iterations in the FindBestClique function, the loop keeps track of best clique so far in terms of the maximum unseen characters covered by the clique.

To prove: After i +1 iterations in the FindBestClique function, the loop keeps track of best clique so far in terms of the maximum unseen characters covered by the clique.

We have proved the correctness of GetUnseenCharsCovered function in the previous claim. So, for the current clique in the i+1 iteration, we know that we will get the unseen\_chars\_covered by it correctly. And using induction hypothesis we know that we have the best clique so far from the previous i iterations. In this iteration, we compare the length of the unseen\_chars\_covered by the $(i+1)^{th}$ clique with the length of the best so far. If the length is greater, we update the best clique so far else we already have the best clique so far. So, we showed that after i +1 iterations in the FindBestClique function, the loop keeps track of best clique so far in terms of the maximum unseen characters covered by the clique.

Using the loop invariant we can show that after len(clique\_list) iterations, the function FindBestClique computes the best clique in terms of the maximum unseen characters covered by a clique.

\textit{\textbf{Claim: }} The function UpdateWordleState updates the state of the wordle tracker correctly.

\textbf{\textit{Proof: }} We already showed in Claim 1 that the state of the wordle game is updated correctly using induction. So we can say that the UpdateWordleState function works correctly for all words of the clique updating the game state accordingly.

\textbf{Time analysis}
Let alphabet size = a

Let there be n words in the vocabulary and the length of each word be l. The worst case number of cliques will be $\binom{N}{k}$ for a fully connected graph if we are finding all k cliques. But the actual number of cliques will be r $<<$ $\binom{N}{k}$.

For each k-clique,

Get unseen chars covered : O(k) because we are performing union operation k times. Then after the loop we are doing intersection operation. Union and intersection can be done in constant time because the alphabet size is fixed "a". So, we can implement the sets using a fixed size boolean array setting True on elements seen and False on elements not seen. So, union operation will be an OR operation across all elements one by one and intersection will be and operation and the cost of both operations will be O(a) = O(1).

Complexity of for loop = $O(\binom{N}{k}k)$

Complexity of UpdateWordleState = $O(k + kl + 1 + nl) = O(kl + nl)$
(complexity of each individual function computed previously)

Total complexity = $O(\binom{N}{k}k + kl + nl)$

\subsubsection{Checking All Anagrams}
\label{attr:check_all_anagrams}

\begin{algorithm}[H]
\caption{checkAllAnagrams}
\label{check_all_anagrams}
\begin{algorithmic}[1]
\Require A Wordle tracker object $WordleTracker$ and the hidden word $HiddenWord$
\Ensure The updated Wordle tracker object $WordleTracker$ after checking all anagrams
\State Initialize an empty dictionary $HiddenWordFict$
\State Initialize an empty dictionary $currGuessDict$
\For{each character $ch$ in $hiddenWord$}
\State Add $ch$ as a key to $hiddenWordFict$ with value as its index
\EndFor
\For{each key $ke$ and value $va$ in $WordleTracker.letterPositions$}
\State Add $ke$ as a key to $CurrGuessDict$ with value as $va$
\EndFor
\While{the length of $wordleTracker.letterPositions$ is less than the length of $hiddenWord$}
\State Get all yellow letters by subtracting the letters found from the letter positions
\State Get all yellow positions by subtracting the letter positions from the range of length of hidden word
\State Initialize $shift$ as 0
\State Set $infoFound$ to False
\While{$infoFound$ is False}
\State Initialize a temporary list $allYellowPositionsTmp$ with the shifted positions of $allYellowPositions$
\State Update the current guess dictionary with the shifted positions
\State Initialize an empty list $guessedWordString$
\For{each key $ch$ and value $pos$ in $currGuessDict$}
\State Set the character at $pos$ in $guessedWordString$ as $ch$
\EndFor
\State Convert $guessedWordString$ to a string
\If{$guessedWordString$ is not in $wordleTracker.wordsGuessed$}
\State Add $guessedWordString$ to $wordleTracker.wordsGuessed$
\EndIf
\For{each true character $trueCh$ and true position $truePos$ in $hiddenWordDict$}
\If{$currGuessDict[trueCh] = truePos$ and $trueCh$ is in $allYellowLetters$}
\State Set $infoFound$ to True
\State Add the true character and its true position to the letter positions dictionary
\EndIf
\EndFor
\State Increment $shift$ by 1
\EndWhile
\EndWhile
\State Initialize an empty list $guessedWordString$
\For{each key $ch$ and value $pos$ in $wordleTracker.letterPositions$}
\State Set the character at $pos$ in $guessedWordString$ as $ch$
\EndFor
\State Convert $guessedWordString$ to a string
\If{$guessedWordString$ is not in $wordleTracker.wordsGuessed$}
\State Add $guessedWordString$ to $wordleTracker.wordsGuessed$
\EndIf
\State \Return $wordleTracker$
\end{algorithmic}
\end{algorithm}

The checkAllAnagrams function takes in a wordleTracker object and a hiddenWord string as input. The purpose of the function is to check all possible anagrams of the letters in hiddenWord and determine which one is the true guess word. The function first creates two dictionaries: hiddenWordDict and currGuessDict. hiddenWordDict maps each letter in hiddenWord to its corresponding index in the word, while currGuessDict maps each letter in the wordleTracker object's letterPositions dictionary to its corresponding position in the current guess word. The function continues until all letters in hiddenWord have been found by creating two lists: allYellowLetters and allYellowPositions. AllYellowLetters contains all the letters that have been found but their positions are not yet known, while allYellowPositions contains all the positions that have not yet been filled with a letter. This has a shifting logic to try out all yellow letters in different positions until some new information is found from them. This shifting logic is also explained at the end.\\
The function then enters another loop that shifts the positions of the letters in allYellowLetters and assigns them to the positions in allYellowPositions. It updates currGuessDict with the new positions of the letters in allYellowLetters. The shifting logic ensures that the function considers all possible combinations of the unknown letters in hiddenWord until the correct guess is found. The function  creates a guessedWordString by filling in the known letters in their correct positions and leaving the unknown letters as underscores. If the guessedWordString has not already been guessed, it is added to the wordleTracker object's wordsGuessed list. It then checks if any of the unknown letters in hiddenWord can be deduced from the current guessedWordString. If so, the letter is added to the wordleTracker object's letterPositions dictionary with its correct position. The loop continues until all letters in hiddenWord have been found. Finally, the function creates a guessedWordString with all the letters in their correct positions and returns the updated wordleTracker object.

\emph{Shifting logic Explanation: }The shifting logic is used to try out different positions for the yellow letters (the unknown letters that have been found but whose positions have not been identified yet) and see if any of them provide new information about the correct positions of the yellow letters. The function starts by finding all the yellow letters and their corresponding yellow positions. It then initializes a variable shift to zero and enters a loop that shifts the yellow positions by one position to the right in each iteration. The letters in currGuessFDict (which represent the guessed positions of the letters found so far) are updated according to the new shifted yellow positions. The shifting logic continues until new information is found or all possible shifts have been tried. The function then returns the updated wordleTracker object.

\subsubsection{\textbf{Time Complexity}}
The time complexity of this algorithm is $O(l)$, where l is the length of the hidden word, assuming comparing guess word with hidden word is a constant time operation. This is because, in the worst case, we rotate by l times, placing each letter at all l positions once, and this gives us information about its correct location. We guess at max l words with each letter's position being shifted by 1 in each of those guess. So, time complexity is $O(l)$.
\subsubsection{Correctness}
\textbf{Claim:} All anagrams are checked given k known letters, and the correct word is guessed.

\textbf{Proof:}
To prove the correctness of the algorithm, we can observe that it follows a simple strategy of finding all possible valid anagrams of the hidden word by iteratively guessing and checking letters in the correct positions.  Overall, this algorithm is guaranteed to find all possible anagrams of the hidden word by systematically trying out all possible combinations of letters and checking for correctness. Only anagrams are checked since the letter values are never changed. Thus, no extra words besides anagrams can be checked. The check of all possible anagrams is ensured by "shifting" the yellow letters in all possible positions until at least one of them turns green.

\subsection{Guessing remaining words}
\label{attr:guess_remaining_words}

\subsubsection{Algorithm description}
In the case that we are not able to find all letters to check for anagrams we have to resort to guessing the remaining words one by one in this method.

The GuessRemainingWords function does this. It in two parameters: wordle\_tracker and the hidden\_word.

For each word in the vocabulary if it has not been discarded, it adds it to the list of guessed words (wordle\_tracker.words\_guessed) and then checks each letter in the word to see if it is a green, yellow, or a grey letter.

If the letter is a green letter (i.e., it appears in the same position in the hidden word), then the letter is added to the set of wordle\_tracker.letters\_found and its position in the hidden word is recorded in the wordle\_tracker.letter\_positions dictionary. If the letter is an incorrect guess (i.e., it does not appear in the hidden word at all), then the letter is added to the set of wordle\_tracker.grey\_letters.

Finally, if all letters in the hidden word have been found, the function calls the CheckAllAnagrams function to check all anagrams of the letters to find the correct answer. Note that the CheckAllAnagrams function simply returns without doing any anything if the current guessed anagram is the actual hidden word, since no yellow letters exist.

\begin{algorithm}
\caption{Guessing remaining words}\label{alg:guess_rem}
\begin{algorithmic}[1]

\Function{GuessRemainingWords}{$wordle\_tracker, hidden\_word$}
    \State $vocab \gets wordle\_tracker.vocab$
    \For {$w \gets vocab$}
        \If {$w \in wordle\_tracker.discarded\_words$}
            \State $continue$
        \EndIf
        \State $wordle\_tracker.UpdateWordsGuessed([w])$
        \State $wordle\_tracker.UpdateLettersFound([w], hidden\_word)$
        \If{$len(wordle\_tracker.letters\_found) == len(hidden\_word)$}
            \Return $CheckAllAnagrams(wordle\_tracker, hidden\_word)$
        \EndIf
    \EndFor
\EndFunction

\end{algorithmic}
\end{algorithm}

\subsubsection{Proof of Correctness}

\textit{\textbf{Claim: }} The function GuessRemainingWords will be able to find the answer (hidden word).

\textit{\textbf{Proof: }} We will use loop invariant to prove this. The loop invariant is  that if the loop runs for i iterations, we haven't found the answer so far.

Base case: Before the first iteration,  this function is called only when all letters in the answer are not found. So, we haven't guessed the hidden word so far. Hence, the loop invariant is true for base case.

Induction hypothesis: If the loop runs for i iterations, we haven't found the answer so far.

To prove: If the loop runs for i+1 iterations, we haven't found the answer so far. We will update the list of words we guessed with the $i+1^{th}$ word. Then we will update the letters found if there are any letters matching with the hidden word. If the length of letters found is equal to the length of the hidden word, means we have found the set of letters in the hidden word. And previously we have shown the correctness of CheckAllAnagrams. And we return from the function when CheckAllAnagrams is called. Any iteration before i+1, we haven't found the word because of induction hypothesis being true.

In the last iteration, just 1 word will be left and that will have to be the answer.

\subsubsection{Time Analysis}
Let the number  of words in the vocabulary be n. Length of each word by l.

In each iteration of the loop:
Updating guessed words : $O(1)$

Update letters found : $O(l)$

Check all anagrams : $O(l)$

Complexity of for loop : $O(l)$

Total complexity of function = $O(nl)$

\clearpage
\subsection{Overall Algorithm Correctness and Time Analysis}
\label{attr:overall_correct_time}

\subsubsection{Proof of Correctness}

\textbf{Claim: } The size of the vocabulary (remaining words = vocab - discarded words) keeps decreasing in each iteration of the outer while loop.

\textbf{Proof: } In each iteration of the while loop, we are first forming a graph using the wordle tracker. We showed that the FormGraph function works correctly. Now only if an edge exists in the graph will we proceed to find cliques in the graph. Now while finding cliques in the graph, if we are able to find even 1 clique we will go ahead and update our wordle tracker. We proved the correctness of the ProcessCliques function. The process cliques function will choose one clique and guess the words part of the clique. So, when words are guessed, the vocabulary will be pruned (i.e. the discarded words set is being updated). There will be only one word in the vocabulary equal to the hidden word. Hence, when a word is guessed it will be either the correct word, or it will give some other pattern (with yellow, grey letters as well) and that pattern will lead to reduction in the size of the vocabulary. 

Now, if the graph is not formed or if we don't find any clique, we break from the loop. In that case we check if we found all letters part of the hidden word, in that case again the vocabulary would have been pruned already by our previous guesses.

If all letters are not found, then we resort to guessing each word one after another, and when we guess each word we will be  discarding words from the vocabulary.

So, in all paths taken by the algorithm we have shown that the size of the vocabulary will keep decreasing.

\textbf{Claim: } The hidden word will always be part of the set remaining words (vocab - discarded words) unless the hidden word has already been guessed.

\textbf{Proof: }  We have previously shown the correctness of FindKClique and ProcessCliques. The while loop in the algorithm is run until we find all the letters of the hidden word. So if all the letters of the hidden word are found, there are two possibilities. Either the hidden word has already been guessed i.e. it was part of the best clique chosen as part of ProcessClique. In that case, we will break out of the loop since the condition for the loop fails. In case the hidden word has not been guessed and we have found all the letters, then we will discard all the words not following the grey/yellow/green pattern of words guessed so far in the ProcessClique sub routine. But the hidden word will not be discarded because the patterns are defined using the hidden word.

If all letters of the hidden word are not yet found, based on the patterns defined by the hidden word, the hidden word will not be discarded and the while loop will continue for further guesses.

\textbf{Claim: } The algorithm will terminate and give the hidden word as the last guess.

\textbf{Proof: } The algorithm continues indefinitely only if in an iteration , the remaining words in the vocabulary remains the same. But we proved previously that the size of vocabulary keeps on decreasing in each iteration of the while loop. We also proved that the hidden word will be present in the remaining set of words as it is always a possible candidate. We will terminate the algorithm when we find all the letters of the hidden word and we proved the correctness of CheckAllAnagrams previously as well. So, we will eventually guess the hidden word correctly.  

\subsubsection{Time Analysis}
 n is the number of words in the vocabulary.

Let alphabet size be a (a=26 in case of letters)

k : $\frac{a}{l}$

Algorithm \ref{alg:play_wordle_clique} (Play Wordle)

Initializing wordle tracker : O(1) 

(Need to setup empty sets and an alphabet sized set of unseen characters.)

Computing time complexity of one iteration of loop : 

FormGraph : $O(n^2)$

FindClique : $O(2^n)$ \\

Since we will be finding k cliques starting from max\_clique\_size till 3 clique (in the worst case).

Process cliques : $O(k\binom{n}{k} + kl + nl)$

Now we need to find the number of iterations of the outermost while loop of Algorithm \ref{alg:play_wordle_clique}(line 5),

The loop runs until all letters in the hidden word are found. So the loop cannot run more than 
"a" times. But we can try to come up with a tighter bound for the number of times the loop runs.

In a single iteration, since we are doing worst case analysis, lets say we are not able to find clique of size max\_clique\_size = $a//word\_length$. Lets we are able to find a clique of size c only (when we don't find a clique of size max\_clique\_size, the max\_clique\_size is reduced by 1 and we find cliques again and repeat this until we find a clique).

So, in the worst case we will find c=3 clique. 

Now, c-clique covers cl letters. So, number of iterations we we are just finding c cliques everytime (since we are doing worst case analysis) = $\frac{a}{cl}$

Therefore total complexity of loop = $O (\frac{a}{c*l}[n^2 + 2^n + k\binom{n}{k} + kl + nl])$

= $O (\frac{a}{c*l}[n^2 + k\binom{n}{k} + 2^n + nl + kl])$ \\

Complexity of CheckAllAnagrams = $O(l)$

Complexity of GuessRemainingWords = $O(nl)$

Either of the function runs. Worst case lets say GuessRemainingWords runs (has the worse time complexity).

Total complexity  = $ O (\frac{a}{c*l}[n^2 + 2^n + k\binom{n}{k} + nl+ kl]+ nl) $

Since, $a,c,l$ are all constants, we can replace them with another constant, and remove them from final time complexity. This gives us,\\
$ O ([n^2 + 2^n + k\binom{n}{k} + nl +  kl] + nl) $

In this term, $2^n$ will dominate the other  terms. This gives us our final overall worst case time complexity - $\boldsymbol{O (2^n)}$

\subsection{Comparing both Approaches}
We compare both the greedy and clique based algorithm using various parameters in the following table:

{\renewcommand{\arraystretch}{1.4}%
\begin{center}
\begin{tabular}{ c|c|c }
 \textbf{Criteria} & \textbf{Greedy} & \textbf{Graph}\\ \hline
\textbf{Easy Mode} & Yes & Yes \\ \hline
\textbf{Hard Mode} & Yes & No \\ \hline
\textbf{Time Complexity} & Polynomial-($O(n^2l^2)$) & Exponential-($O (2^n)$) \\ \hline
\end{tabular}
\end{center}
}
In the case of the graph algorithm, we aim to first find all the $k$ alphabets present in the hidden word. Our subsequent guesses aim to find the unknown alphabet and hence we do not constrain the algorithm to repeat any yellow or green alphabets found in previous rounds. This makes the graph algorithm work only in the easy mode. \\
On the other hand, in the case of the greedy algorithm, at each round, our vocabulary is pruned such that if a yellow letter is present in the previous round, it has to be present in all words of the pruned vocabulary. Similarly, if a green letter is present in the previous round, it has to be present in the same location as all the words in the pruned vocabulary. Our guess is made only from the pruned vocabulary and hence satisfies the constraints of the hard mode. \\
On comparing the time complexities, we see that we get polynomial time complexity for the greedy algorithm and an exponential time complexity for the clique based algorithm.
\section{Experiments}
\subsection{Proposed Simulations}
We propose to implement the wordle game, both the algorithms and do the following simulations.
\begin{itemize}
    \item Play the game for vocabulary length of $l=3,4,..,9$ using all words in the vocabulary as hidden words 
    \item Compute the average number of guesses it takes to find the hidden word for both the algorithms, for all possible hidden words, for all possible values of $l$
    \item For the greedy algorithm, find the best-starting word for each $l=3,4,..,9$ 
    \item Vary the number of guesses $m$, and see the win percentage $\%$ for each combination of $l$ and $m$
    \item For the greedy algorithm, find the word which takes the maximum number of guesses
    \item For the graph algorithm, find the largest clique that is found for every $l$
    \item The graph algorithm approach is exponential in the clique formation step, so we analyse its runtime with respect to different values of k.
\end{itemize}
\subsection{Greedy Algorithm}
In this analysis section, we explore the performance of a greedy algorithm on the Wordle game. Specifically, we investigate how the performance of the algorithm changes as we increase the length of the word and the size of the word set. We conduct five experiments using different word set sizes with varying lengths of words, ranging from 3 to 9 letters. For each experiment, we measure the average number of guesses required to solve the game using the greedy algorithm with a maximum number of tries allowed. Additionally, we analyze the win percentage as we increase the number of tries allowed, and identify the worst hidden words present in the vocabulary that take the most tries to guess. The results of these experiments shed light on the performance of the greedy algorithm on Wordle and provide insights into the difficulty of the game for longer words and larger word sets.

\subsubsection{Experiment 1: The best first word to guess based on the vocabulary}

Since, the greedy algorithm takes in all possible combinations of hidden words and sticks to predicting the word that decreases the worst-case tries, the first word predicted by the algorithm for a k-letter game remains constant for all hidden words. Hence, for each of the k-letter games possible ($k \in [3,9]$), we list the first word that our greedy algorithm guesses in Table \ref{tab:best_first}.

\begin{table}[h]
\centering
\begin{tabular}{|c|c|}
\hline
\textbf{$k$-letter} & \textbf{Word set size} \\
\hline
3 & 1176 \\
4 & 3836 \\
5 & 6373 \\
6 & 8077 \\
7 & 7882 \\
8 & 6074 \\
9 & 3766 \\
\hline
\end{tabular}
\caption{Number of words present in the vocabulary for each word size-$k$}
\label{tab:word_size}
\end{table}

One of the reasons that the words listed in Table \ref{tab:best_first} are the best first words to guess for their respective k-letter games, is because they contains a good mix of letters that are commonly used in English and these words help maximize the potential of guessing other words.
\\
\begin{table}[h]
\centering
\begin{tabular}{|c|c|}
\hline
\textbf{$k$} & \textbf{Value} \\
\hline
3 & iao \\
4 & eoan \\
5 & nares \\
6 & tanier \\
7 & latrine \\
8 & corsaint \\
9 & columnist \\
\hline
\end{tabular}
\caption{Best first guess for specified values of $k$}
\label{tab:best_first}
\end{table}

\subsubsection{Experiment 3: Optimizing on time by fixing first guess}

Through our experiments and time complexity derivation, we found that guessing the first optimal word according to our greedy algorithm takes the highest amount of time. This occurs mainly due to the fact that our vocabulary set is untrimmed and our greedy algorithm runs in $O(n^{2}l^{2})$, where n is the length of the vocab. Hence, the larger the vocab the greater the number of comparisons made during the algorithmic run.\\
So we perform two experiments. In the first experiment, we run the greedy algorithm for all words and all lengths and see the time taken. In the second experiment, we run the optimized version of the algorithm where we store the first guessed word for each length and use it to guess the subsequent words. The optimized version returns the same results as the original version since the first guess word remains the same for a fixed length of words in the vocabulary. We see a drastic improvement in the run time as seen in the table below.

\begin{table}[ht]
\centering
\begin{tabular}{|c|c|c|c|}
\hline
\textbf{$k$-letter} & \textbf{First guess time} & \textbf{GreedyAlgo} & \textbf{Optimized GreedyAlgo} \\
\hline
3 & 1s & $\sim19m$ & 68s \\
4 & 12s & $\sim13h$ & 614s \\
5 & 36s & $\sim65h$ & 483s \\
6 & 67s & $\sim149h$ & 275s \\
7 & 71s & $\sim155h$ & 99s \\
8 & 47s & $\sim79h$ & 59s \\
9 & 20s & $\sim21h$ & 59s \\
\hline
\end{tabular}
\caption{Time analysis for Greedy algorithm with and without optimization}
\end{table}

\subsubsection{Experiment 2: Average number of guesses}

We found that as the number of letters in the word increases, the average number of guesses decreases. This indicates that longer words are easier to guess than shorter words. One of the reasons for this trend is that longer words take a lesser number of tries to explore all the 26 letters of the English language. Moreover, the nature of the game itself plays a crucial role in the average number of guesses required to solve the game. For shorter words, it is more likely that the guesses will be spread out among many different letters, resulting in a higher number of guesses required to solve the word. On the other hand, for longer words, the number of possible words to guess from is limited and more likely to be concentrated around the few remaining letters, making it easier to guess the word in fewer attempts. This trend is observed in the table, where the average number of guesses decreases as the number of letters in the word increases.

\begin{table}[h]
\centering
\begin{tabular}{|c|c|}
\hline
\textbf{$k$-letter} & \textbf{Average tries} \\
\hline
3 & 8.66 \\
4 & 6.49 \\
5 & 4.84 \\
6 & 4.23 \\
7 & 3.65 \\
8 & 3.14 \\
9 & 2.8 \\
\hline
\end{tabular}
\caption{Average number of guesses required for specified values of $k$}
\end{table}

\subsubsection{Experiment 3: Win percentage chart}

\begin{figure}[h]
    \centering
    \includegraphics[width=0.8\textwidth]{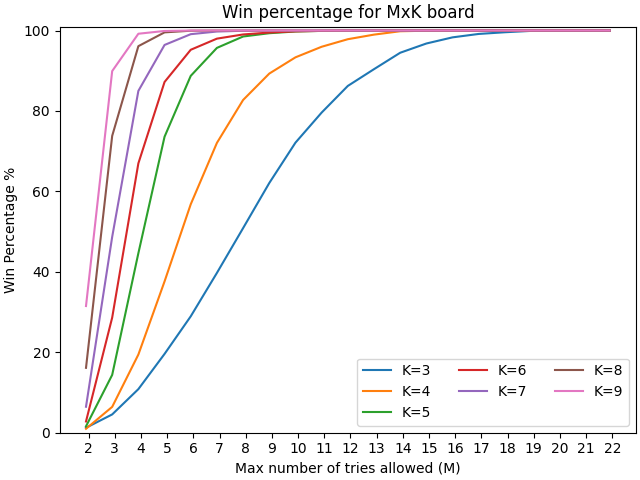}
    \caption{Winning percentage using the Greedy algorithm for different values of $m$ and $k$.}
    \label{fig:winning_percentage}
\end{figure}

In this experiment, we aimed to analyze the effectiveness of our proposed Greedy algorithm by creating a win percentage chart for different numbers of allowed tries. As expected, we observed that as the number of tries increases, the win percentage also increases. The results of our simulation are shown in Figure \ref{fig:winning_percentage}

Upon analyzing the chart, we made several interesting observations. Firstly, we noticed that for all k-letter games where k $\in$ [3,9], our proposed Greedy algorithm was able to win within 22 tries. Additionally, we observed that for all k-letter Wordle games, the win percentage is above 90\% within 13 tries. These findings suggest that our algorithm performs well across all word set sizes and is able to achieve high win rates with relatively few tries.

Furthermore, we found that for the standard Wordle game with $m=6$ and $k=5$, our algorithm had a win percentage of 88.72 for a word set size of 6373. This indicates that our algorithm is effective for the most commonly played Wordle game.

Finally, we found that for the 5-letter game, our Greedy algorithm is able to win every single game if we increase the maximum number of tries to 13. This highlights the potential of our algorithm to perform even better with slight modifications in the number of allowed tries.

\subsubsection{Experiment 4: Worst hidden words}

We identified the worst hidden words present in the vocabulary that take the most tries. Table \ref{tab:table_worst} shows the worst hidden words for each k-letter word:

\begin{table}[ht]
\centering
\begin{tabular}{|c|c|c|c|c|c|}
\hline
\textbf{$k$-letter} & \textbf{Worst word 1} & \textbf{Worst word 2} & \textbf{Worst word 3} & \textbf{Worst word 4} & \textbf{Tries} \\
\hline
3 & uta & - & - & - & 22\\
4 & zain & - & - & - & 17\\
5 & mesal & - & - & - & 14\\
6 & ostrea & - & - & - & 14\\
7 & vinegar & - & - & - & 10\\
8 & jitendra & overstay & ringdove & - & 7\\
9 & hairstone & strobilae & swanimote & tribesman & 6\\
\hline
\end{tabular}
\caption{Worst hidden words to guess for specified values of $k$}\label{tab:table_worst}
\end{table}
\subsection{Clique Based Algorithm}

We choose a different vocabulary set for our Clique experiments than the set used above in the Greedy approach. This is because we desire more words in the vocabulary, which will help us find cliques if they exist. This is more critical for cases like finding 5 and 6 cliques where cliques are rare. Therefore, our vocabulary has more number of words than the greedy experiments.
Time analysis to compute the cliques for all l letters words. l = 4 to l=8. For each l letter set of words, we see how much time it takes to compute all cliques.

\begin{table}[h]
\centering
\begin{tabular}{|c|c|}
\hline
\textbf{$l-letter$} & \textbf{Number of words} \\
\hline
4 & 5549 \\
5 & 10175 \\
6 & 13857 \\
7 & 13661 \\
8 & 10428\\
\hline
\end{tabular}
\caption{Number of words for each l in the vocabulary}
\end{table}

We can see that clique formation time exponentially increases as the $l$, ie , number of letters in the word increase. One interesting observation is the spike in clique formation time between $l=5$ and $l=6$. In this case, we noticed that the graph itself was way more dense for $l=5$ than $l=6$ because there are more number of words with 0 overlapping letters in case of 5 letter words, than 6 letter words. Since the complexity of clique formation was $\binom{N}{k}$, where $k=\frac{ALPHABET\_SIZE}{l}$, there is an interplay of number of words, and the number of letters. This is best visible in the case between $l=5$, and $l=6$. Here, for $l=5$, both the clique size and the number of edges is larger than $l=6$ which leads to the huge increase in runtime for $l=5$.

\begin{table}[h]
\centering
\begin{tabular}{|c|c|c|c|}
\hline
\textbf{$l-letter$} & \textbf{Number of Vertices} & \textbf{Number of edges} & \textbf{Clique size $k$ = $26/l$}  \\
\hline
4 & 5549 & 5267897 & 6  \\ 
5 & 10175 & 7641781 & 5 \\  
6 & 13857 & 4415108 & 4 \\ 
7 & 13661 & 874963 & 3  \\ 
8 & 10428 & 53013 & 3 \\
\hline
\end{tabular}
\caption{Number of vertices and edges in each graph that is trying to find clique size $k$ for $l$ letters.}
\end{table}

\begin{table}[h]
\centering
\begin{tabular}{|c|c|c|}
\hline
\textbf{$l-letter$} & \textbf{Time taken} & \textbf{Number of $k= 26/l$ cliques found} \\
\hline
4 &  1700+ hours(est) & -\\ 
5 & 14+ hours(est)& - \\  
6 & 841.06s & 5  \\ 
7 & 81.93s & 1877 \\ 
8 & 42.14s & 0 \\
\hline
\end{tabular}
\caption{Number of $k$ cliques found for l letter words.}
\end{table}

\begin{figure}[h]
    \centering
    \includegraphics[width=0.8\textwidth]{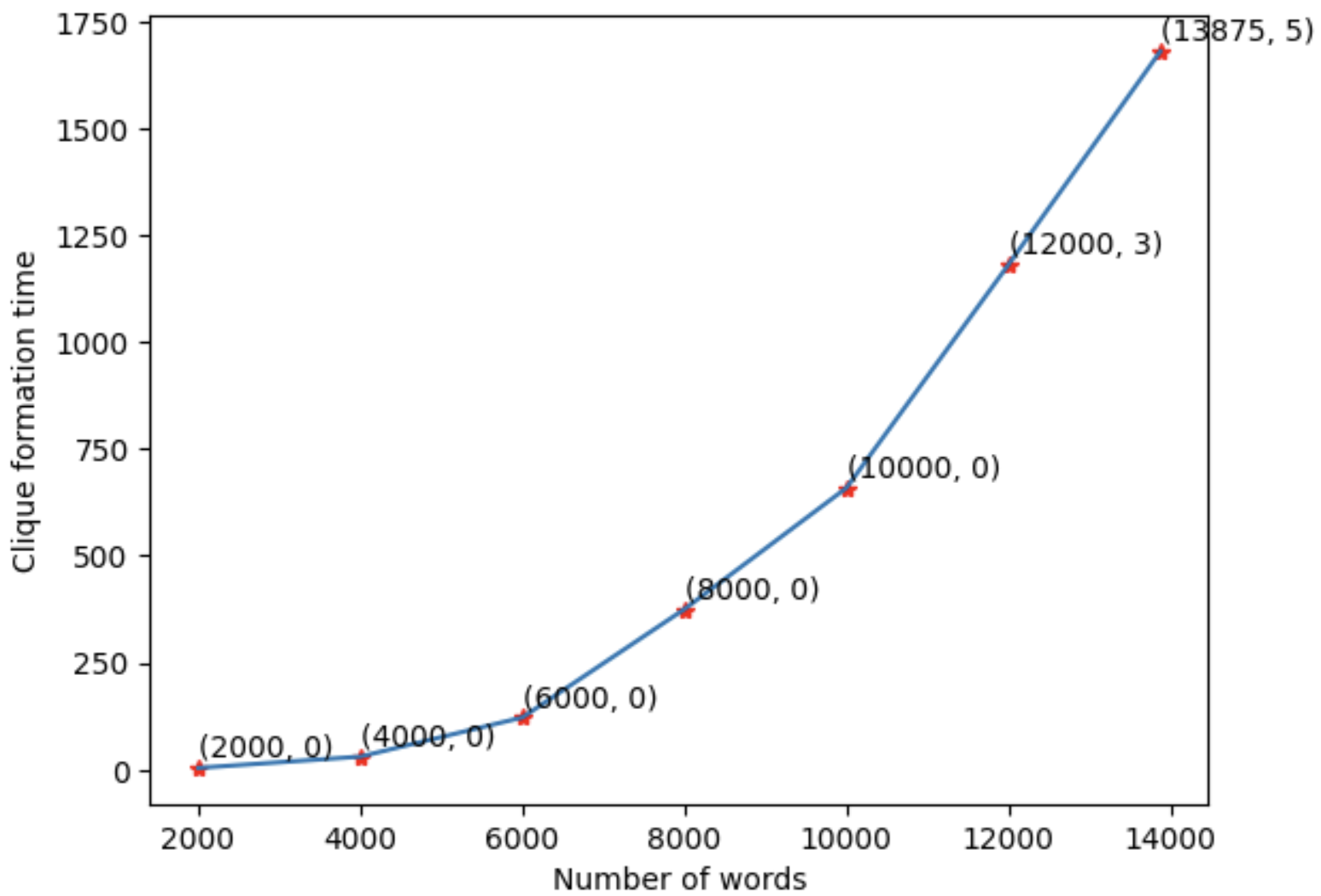}
    \caption{Plot of Clique formation time vs Number of Words (along with number of cliques found). A subset of total words taken for each run} 
    \label{fig:cliq}
\end{figure}

\newpage We further analyse for the case of $l=6$ how clique formation time and result changes with the change in vocabulary size as shown in Figure \ref{fig:cliq}. We see that with increase in vocabulary size (taking a subset of all words), runtime increases drastically, but also that large vocbulary is necessary to find cliques. We also show some of the cliques that are found. We choose $l=6$ because it runs in a desirable duration of time (unlike for $l=5$) and also highlights the need for more vocabulary size to discover cliques ( unlike for $l=7$ which has high number of cliques)

\begin{figure}[h]
    \centering
    \includegraphics[width=0.8\textwidth]{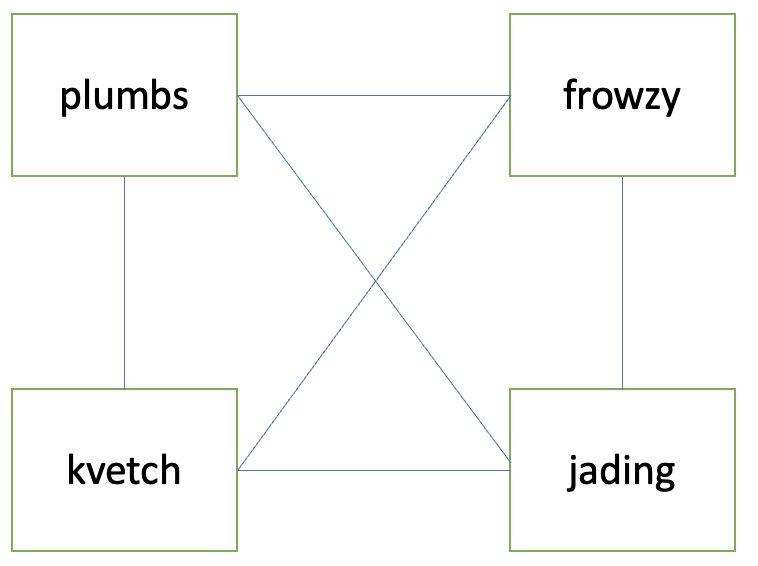}
    \caption{Example 4-clique for 6 letter words}
    \label{fig:cliq4}
\end{figure}

We show a clique we obtained for l=6 case. We create a 4 clique ($26//6$). We can see in Figure \ref{fig:cliq4} that there are edges between two words only when they don't have any letter in common between them. We can have 4 guesses using this clique and we cover 24 letters.

\newpage
\section{Conclusion}

On comparing two methods, we see that the greedy algorithm is more robust in terms of running both the easy and hard modes while the clique algorithm can only run the easy mode. The clique 
 based algorithm is exponential in runtime because of which it is difficult to obtain results while the greedy algorithm has polynomial time complexity. 

 We have presented the results for both algorithms, and potential future direction for these algorithms can be extending the algorithm to words containing repeated letters, and further optimizing the clique finding method. Also, other algorithmic paradigms can be tried to solve the Wordle game.

\printbibliography

\end{document}